\documentclass[twocolumn]{revtex4-1} 
\usepackage{geometry}                
\usepackage{graphicx}
\usepackage{amssymb}
\usepackage{stfloats}
\usepackage{xspace}
\usepackage{url}
\usepackage{float}
\usepackage{multirow}
\usepackage{booktabs}
\usepackage{graphics}
\include{00README.XXX}

\begin{document}
\title{Lifetime measurement of the $^{229}$Th nuclear isomer}

\author{Benedict Seiferle}
\affiliation{Ludwig-Maximilians-Universit\"at M\"unchen, Am Coulombwall 1, 85748 Garching b. M\"unchen, Germany }
\author{Lars von der Wense}
\affiliation{Ludwig-Maximilians-Universit\"at M\"unchen, Am Coulombwall 1, 85748 Garching b. M\"unchen, Germany }
 \author{Peter G. Thirolf}
 \affiliation{Ludwig-Maximilians-Universit\"at M\"unchen, Am Coulombwall 1, 85748 Garching b. M\"unchen, Germany }

\abstract{%
The first excited isomeric state of $^{229}$Th possesses the lowest energy among all known excited nuclear states.
The expected energy is accessible with today's laser technology and in principle allows for a direct optical laser excitation of the nucleus.
The isomer decays via three channels to its ground-state (internal conversion, $\gamma$ decay and bound internal conversion), whose strengths depend on the charge state of $^{229m}$Th.
We report on the measurement of the internal-conversion decay half-life of neutral $^{229m}$Th. 
A half-life of 7 $\pm$ 1 $\mu$s has been measured, which is in the range of theoretical predictions and, based on the theoretically expected lifetime of $\approx10^4$ s of the photonic decay channel, gives further support for an internal conversion coefficient of $\approx10^9$, thus constraining the strength of a radiative branch in the presence of IC.
} 
} 
\maketitle 
$^{229}$Th is the only known nucleus providing an excited isomeric state of sufficiently low energy to allow for direct nuclear optical laser excitation \cite{NNDC}.
The possibility to drive the transition with laser technology has led to the proposal of a multitude of interesting applications.
The predicted spectroscopic properties of the $^{229m}$Th ground-state transition make it a promising candidate for a nuclear optical clock that may outperform today's existing atomic clock technology \cite{Peik1,CampbellClock, Rellergert}. 
As other ultra-precise optical clocks, a nuclear clock could be a tool in the search for dark matter \cite{DarkMatter}, gravitational waves \cite{GravitationalWaves} as well as for geodesy \cite{Geodesy}.
Such a nuclear clock promises ultra-high sensitivity for potential time variations of fundamental constants \cite{Flammbaum}.
There exist also proposals towards a nuclear $\gamma$-ray laser \cite{GammaLaser} based on the $^{229m}$Th ground-state transition and a nuclear qubit for quantum computing \cite{QuantumComputing}.
However, to enable laser excitation, precise knowledge on the spectroscopic properties of the transition, such as the lifetime and the excitation energy, is required.
Since the first proposal of the existence of a low-energy isomeric state of $^{229}$Th in 1976 \cite{1976} several indirect energy measurements \cite{1990,1994,Becketal} have been performed.
With steadily improved detector energy resolution, these measurements pinned down the energy to 7.8 $\pm$ 0.5 eV \cite{7.8eV} ($\lambda\approx$ 159 $\pm$ 10 nm).
A direct half-life measurement of a photonic decay channel \cite{6+-1} has been controversially discussed \cite{PeikComment}.
Several different experimental approaches have been pursued, aiming for a measurement of the isomer's properties or for an optical laser excitation of the nucleus \cite{Jeet, Stellmer, Porsev, PTB, Kazakov, Campbell, Lars-Jinst, Ian, EPJD, Swanberg}. 
However, despite significant experimental effort, no conclusive measurement of the isomeric half-life has been reported so far.
\\
\begin{figure*}[ht]
\includegraphics[width = 0.9\textwidth]{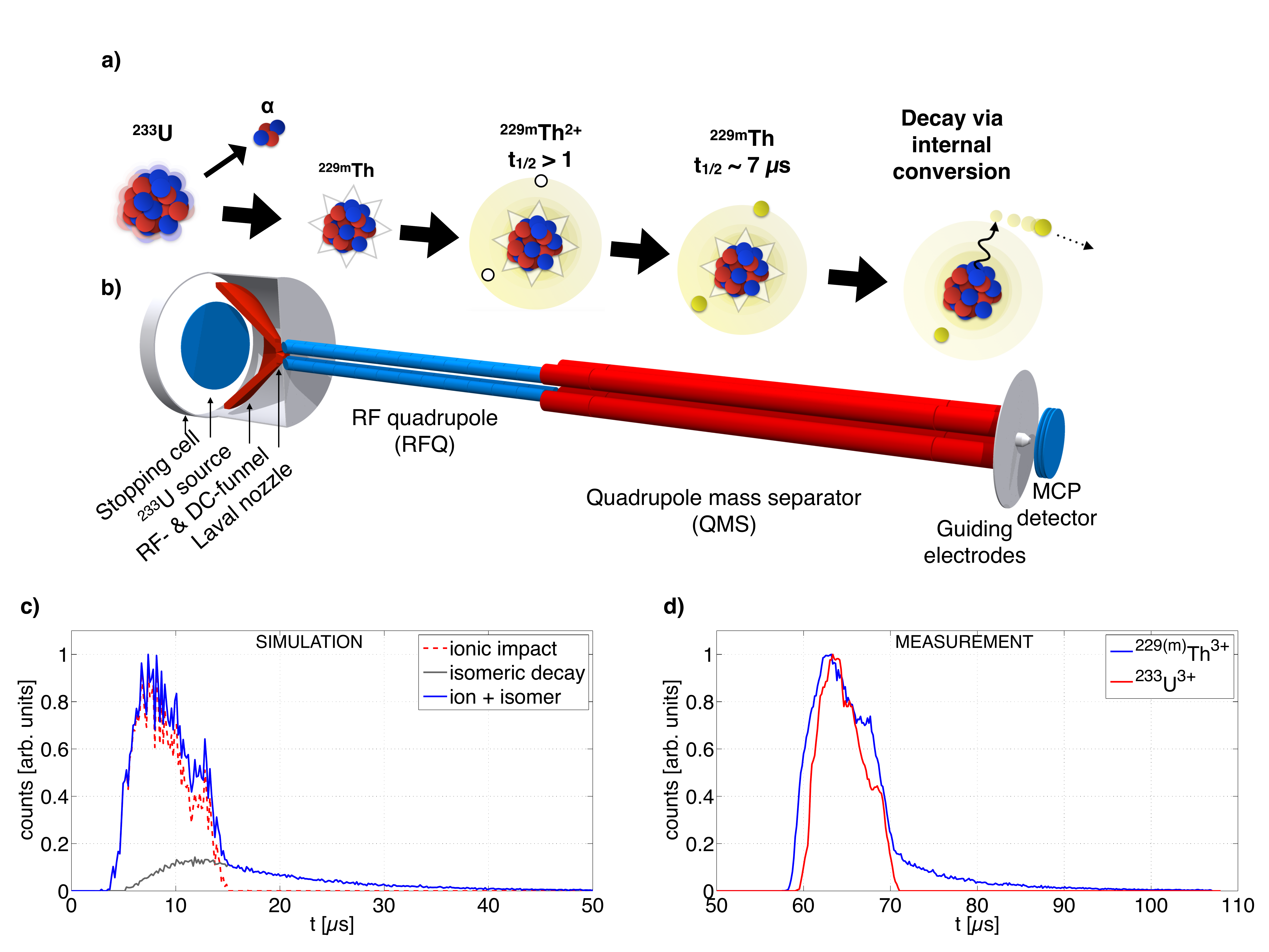}
\caption{ \textbf{a)} Visualization of the detection scheme reported in Ref.~\cite{LARS}. %
An ion beam is formed from $^{229(m)}$Th ions emerging from a $^{233}$U $\alpha$-recoil source. %
As long as $^{229m}$Th remains in a charged state, the lifetime is longer than 1 minute. %
In the moment when the ions neutralize (\textit{e.g.} by collecting the ions on a metal surface), internal conversion is triggered and the lifetime will be in the range of $\mu$-seconds. 
The emitted electron can be detected.
 \textbf{b)} Scheme of the experimental setup that was used. A detailed description is given in the text.
 \textbf{c)} Simulation of the isomer decay time characteristics of $^{229}$Th bunches. 
 The simulation is based on a measured bunch shape, the 2\% of the $^{229}$Th ions that, according to the nuclear level scheme, are in the isomeric state and an assumed half-life of 7 $\mu$s after neutralization. 
 The electron detection efficiency is assumed to be 25 times larger than the one for ion detection.
\textbf{d)} Corresponding measurement of the isomeric decay with a bunched $^{229(m)}$Th$^{3+}$ ion beam (blue) and a comparative measurement with $^{233}$U$^{3+}$ (red). 
\label{ExpScheme}}
\end{figure*}
As a result of the low excitation energy, the isomer can decay via three decay channels to its ground state, whose occurrence depends on the electronic surrounding of the nucleus \cite{Strizhov,ElectronicEnvironment, ElectronicTkalya,ElectronicEnvironment2}:
when the binding energy of an electron $E_B$ in the surrounding of the nucleus is lower than the excitation energy of the isomer $E_I$,  the isomer decays preferably via internal conversion (IC) by emitting an electron with an energy $E_{e^-} = E_{I} - E_{B}$. 
For the expected isomer energy of 7.8~eV, this channel dominates for neutral $^{229}$Th, which has an ionization energy of 6.3 eV \cite{Ionization}.
Other decay channels of the isomeric state are (i) via the emission of a $\gamma$ ray or (ii) via bound-internal conversion (BIC), where an electronic shell state is excited. 
The different decay channels lead to significantly different isomeric lifetimes: the isomer in charged $^{229}$Th$^{n+}$, where IC is energetically forbidden, is expected to be up to 10$^9$ times longer-lived (with a lifetime ranging from minutes to hours) than in case of neutral isolated $^{229}$Th \cite{TkalyaJeet}.

A recent direct measurement of the isomer's internal conversion decay channel provided a detection scheme for the identification of the isomeric properties \cite{LARS}.
In this letter, we adapt this detection scheme and report on the measurement of the lifetime of the $^{229m}$Th internal-conversion decay.
A lifetime of 7$\pm$1 $\mu$s has been inferred, which is in agreement with theoretical predictions, in this way supporting the potential use of $^{229m}$Th$^{n+}$ as a highly stable nuclear frequency standard.
The approach reported in Ref.~\cite{LARS} makes use of the isomer's lifetime dependency on the charge state and is visualized in Fig.~\ref{ExpScheme}a:
$^{229m}$Th is populated via the 2\% decay branch in the $^{233}$U $\alpha$-decay \cite{Barci}.
Emerging from a thin $^{233}$U layer, $^{229(m)}$Th $\alpha$-recoil ions remain in a charged state, thus IC is forbidden and the half-life is expected to be in the range of minutes to hours.
An ion beam can be formed (within several milliseconds), which is then accumulated directly on a metal surface.
As soon as the $^{229(m)}$Th ions neutralize, IC is triggered and an electron is emitted that is measured with a multi-channel plate (MCP) detector.
In our measurements, the ions are directly collected on the surface of the MCP detector, that is used for both, the charge capture of the ion and the detection of the IC electron.\\
The setup is schematically shown in Fig.\ref{ExpScheme}b. 
It consists of a $^{233}$U $\alpha$-recoil source (290 kBq, electro-deposited with $\o$=90 mm on a titanium-sputtered silicon wafer), which is placed in a buffer-gas stopping cell, filled with 40 mbar of ultra-pure helium. 
A DC offset of 70 V is applied to the source.
The recoil ions are stopped by the buffer gas and are guided by 50 ring electrodes, arranged in a funnel-like structure (RF- \& DC-funnel in Fig.~\ref{ExpScheme}b).
RF voltages (V$_{\mbox{\scriptsize{pp}}}$= 120 V at 850 kHz) and a DC gradient (ranging from 65 V to 33 V) are applied to the electrode structure.
In this way, the ions are guided towards a Laval nozzle ($\O$ 0.6~mm nozzle throat) that connects the stopping cell to the subsequent extraction vacuum chamber (p~$\approx$~10$^{-4}$~mbar), housing a radio-frequency quadrupole (RFQ).
The thereby created supersonic helium gas jet drags the ions off the field lines and injects them into the RFQ. 
The RFQ consists of twelve segments, to which an RF voltage (V$_{\mbox{\scriptsize{pp}}}$= 200 V at 880 kHz) is applied. 
Each segment can be set to an individual DC offset. 
Typically, a voltage gradient of $-$0.1 V/cm is applied to the first nine segments, ranging from 32.0 V to 30.2 V.
The 10$^{\mbox{\scriptsize th}}$ and the 11$^{\mbox{\scriptsize th}}$ segment are set to 28.0 V and 25.0 V, respectively.
The DC offset of the 12$^{\mbox{\scriptsize th}}$ and last segment is switchable within 0.1 $\mu$s with a fast solid-state high-voltage switch between 34.0 V and 0 V. 
Thereby it is possible to create a potential well to trap and cool the ions in the region of the 11$^{\mbox{\scriptsize th}}$ segment, then release them within a fraction of a $\mu$s and thus create a short ion bunch. 
The ions are loaded into the trap for 80 ms.
While the ions are cooled in the trap (10 ms) and released, the source is set to a DC offset of 0 V.
At this point still all daughter nuclei are contained in the ion bunch. 
To select only ions with a specific mass-to-charge ratio, a quadrupole mass separator (QMS) (built according to \cite{Haettner}) is installed behind the RFQ bunching unit in a separately pumped vacuum chamber (p~$\approx$~10$^{-6}$~mbar).
The mass resolution of the QMS was found to be $m/\Delta m = 150$ at 70\% transmission efficiency \cite{Lars-EPJA}.
Following the QMS, there are three ring electrodes that guide the ion beam towards a detector (guiding electrodes in Fig.\ref{ExpScheme}b). 
The efficiency of the entire setup in continuous extraction mode is discussed in detail in Ref.~\onlinecite{Lars-EPJA}.
The ion bunches used in our measurements exhibit an ion time-of-flight (TOF) width of about 10~$\mu$s with a falling edge of $<$~1~$\mu$s (see Fig. \ref{ExpScheme}d)).
One bunch of $^{229(m)}$Th ions in the 2+ or 3+ charge state contains about 320 and 400 ions, respectively.
The MCP detector that is used to collect the ions and measure the IC electrons is a two-stage MCP in chevron geometry (Hamamatsu Type F-2223, 25 mm diameter active surface) with a nickel alloy surface. 
No special cleaning procedure was applied to the MCP surface prior to ion collection.
The ions are collected with low kinetic energy in order to reduce  parasitic signals originating from ionic impact. 
The MCP was operated at low negative surface voltage (typically at $-$70 V).
A potential difference of +1700 V between the MCP surface and the second stage and +200 V difference between the second stage and the MCP anode was applied.
The MCP signal is pre-amplified (ORTEC VT-120 Fast Timing Preamplifier) and recorded with a multichannel scaler (Stanford Research SR-430), typically in bins of 160 ns. 
The falling edge of the switchable last RFQ segment is fed to a constant fraction discriminator and subsequently used as a start signal for the buncher.
\begin{figure*}[ht]
\includegraphics[width = 1\textwidth]{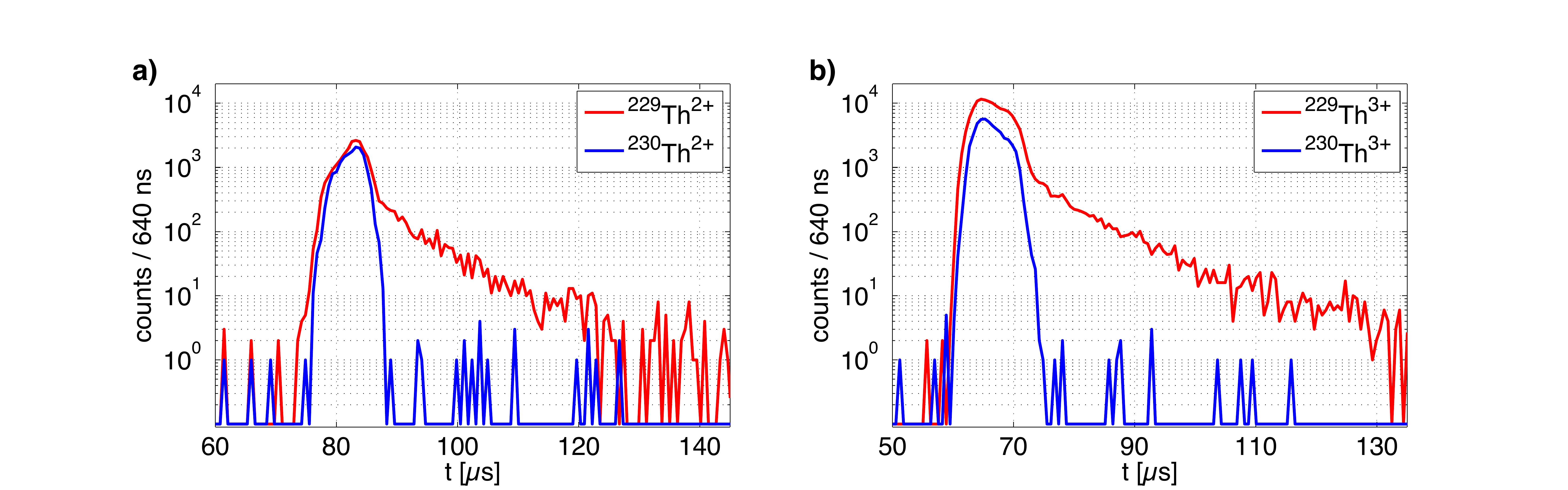}
\caption{%
Measurement of the isomeric decay with a bunched \textbf{a)} $^{229(m)}$Th$^{2+}$ and \textbf{b)}$^{229(m)}$Th$^{3+}$ ion beam (red).
Corresponding comparative measurements performed with $^{230}$Th$^{2+}$ and $^{230}$Th$^{3+}$ are also shown (blue). Note that for all the logarithmic plots above bins containing 0 counts were subsequently set to 0.1 counts in order to visualize the baseline. \label{Th230}}
\end{figure*}
\begin{figure}[ht]
\includegraphics[width = 0.5\textwidth]{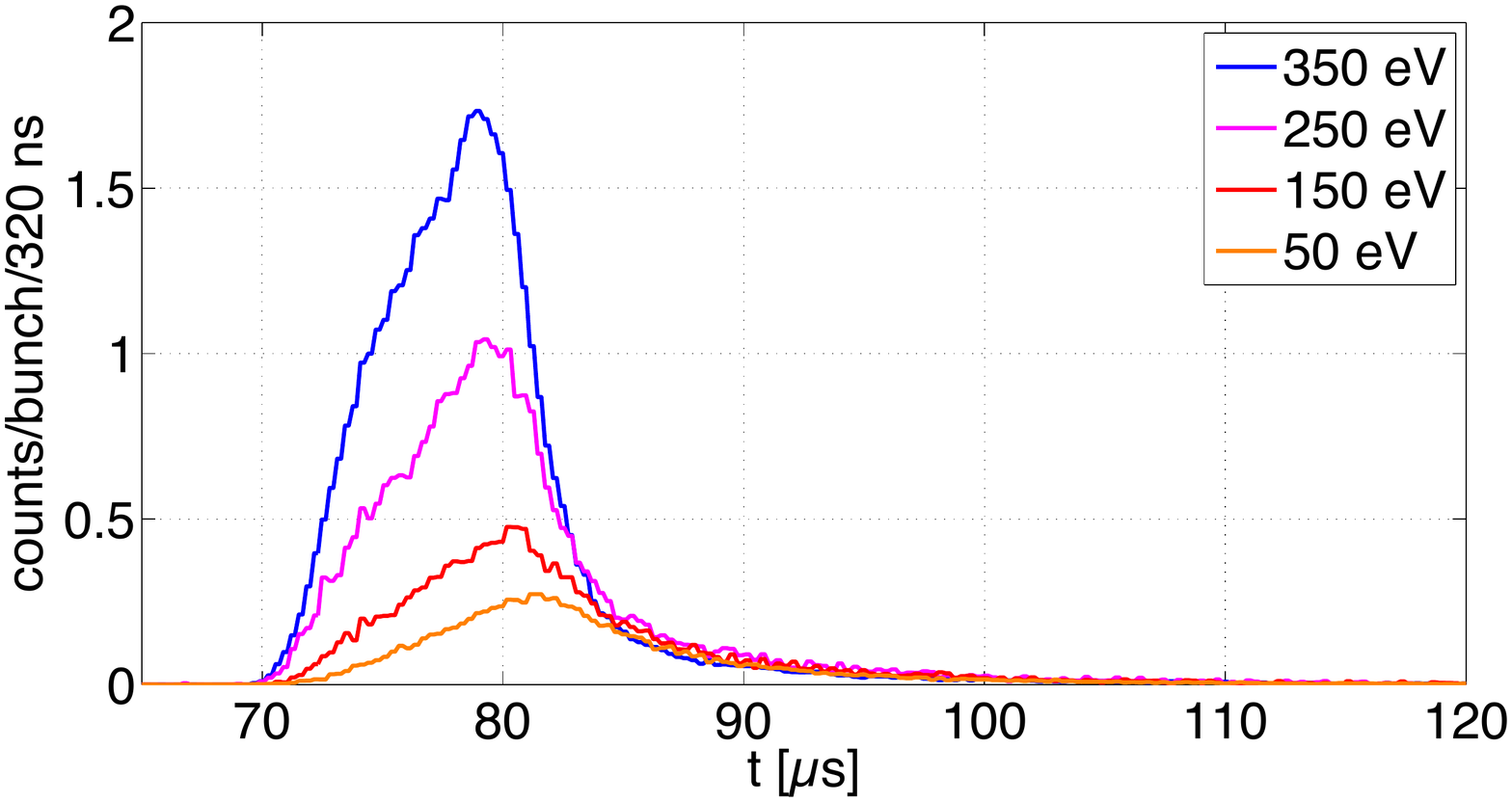}
\caption{Signal obtained when collecting $^{229(m)}$Th$^{2+}$ ion bunches with different kinetic energies.
The isomeric decay signal strength remains constant in all measurements, while the ionic impact signal decreases with the kinetic energy of the ion bunch. \label{Gradient}}
\end{figure}
Figure \ref{ExpScheme}c shows a simulation of the expected isomer decay time characteristics for a bunched ion accumulation on an MCP detector.
In the simulation it is assumed that 2\% of the ions are in the isomeric state and that the detection efficiency ratio between low energy electrons ($\epsilon_{e^-}$) and ions ($\epsilon_{ions}$) is $\epsilon_{e^-}/\epsilon_{ions} = 25$. 
The signals from several bunches recorded with $^{233}$U$^{3+}$ ions were used as an input for the bunch shape and the isomeric decay time characteristics was simulated with a Monte-Carlo based code.
For the simulations a half-life of 7 $\mu$s was assumed.
The grey curve in Fig.~\ref{ExpScheme}c represents the isomeric decay signal, the red curve the ionic impact signal and the blue curve shows the total signal that will be obtained in a measurement.
It is clearly visible that an isomeric decay signal extends beyond the ionic impact signal caused by the accumulation of the ion bunch.
Figure \ref{ExpScheme}d shows the corresponding measurements performed with $^{229(m)}$Th ions (blue) and $^{233}$U ions (red) in the 3+ charged state, collected on the MCP detector at -70 V surface voltage. 
As in the simulations, an exponential decay after the ion accumulation occurs in the measurements performed with $^{229(m)}$Th, which is attributed to the isomeric decay of $^{229m}$Th.
The shift of $\approx$1 $\mu$s in the time-of-flight signal of ionic impact between $^{229(m)}$Th$^{3+}$ and $^{233}$U$^{3+}$ is due to the mass difference of $^{229(m)}$Th and $^{233}$U.

In order to exclude chemical effects on the surface as a signal origin, comparative measurements with $^{230}$Th ions have been conducted.
$^{230}$Th ions are extracted from a $^{234}$U $\alpha$-recoil source with an activity of 270 kBq and identical geometric dimensions as for the $^{233}$U $\alpha$-recoil source.
The data for $^{229(m)}$Th ions (red curves) and $^{230}$Th ions (blue curves) in the 2+ and 3+ charged state, collected on the MCP detector at -100 V surface voltage, is plotted in Fig.~\ref{Th230}.
Both measurements were performed under comparable conditions (20,000 bunches for $^{229(m)}$Th and 10,000 bunches for $^{230}$Th, MCP surface voltage: $-100$ V) and show the ionic impact of the $^{229(m)}$Th and $^{230}$Th ions.
Only for $^{229(m)}$Th ions a decay signal remains after the ionic impact, in this way excluding any atomic-shell based effects as signal origin.
Therefore the decay can be attributed to the $^{229m}$Th isomer to ground-state transition via internal conversion.

\begin{figure*}
\includegraphics[width = 1.2\textwidth]{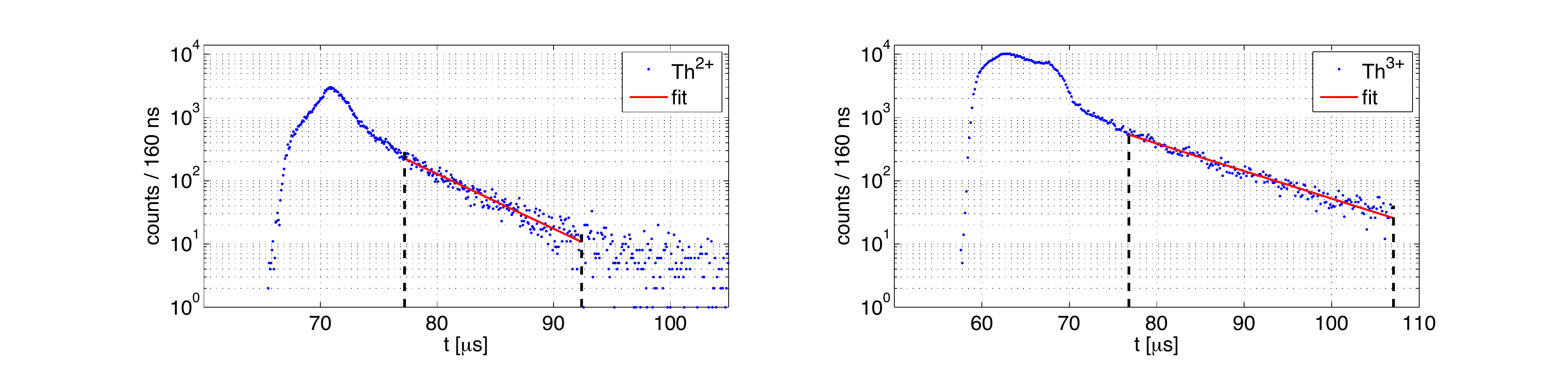}
\caption{Logarithmic plot of the temporal decay characteristics for $^{229(m)}$Th$^{2+}$ (left) and $^{229(m)}$Th$^{3+}$ ions (right) together with the fit curves (red) applied to determine the isomeric half-life of $^{229m}$Th after charge recombination on the MCP detector surface. The fit-range is indicated by the vertical black dashed lines.\label{Analysis}}
\end{figure*}
For a further distinction between the ionic impact signal and the isomeric decay signal, their behavior at different kinetic energies of the ions is investigated. %
A measurement series with different MCP surface voltages $U_S$ was performed, which is shown in Fig. \ref{Gradient}.
For this measurement series, the QMS was set to select only $^{229(m)}$Th$^{2+}$ ions. 
The 2+ charged state is chosen, since here lower kinetic energies can be achieved. 
The plotted curves show the number of counts per bunch.
When the collection voltage is reduced, the strength of the tail behind the ionic impact signal stays constant, while the ionic impact signal decreases, due to the reduced ion detection efficiency of the MCP at lower kinetic energies.

Measurements with $^{229(m)}$Th$^{1+}$ have been performed, but no signal that could be attributed to the isomeric decay has been observed. 
Although the singly charged state is not extracted efficiently in this experimental setup \cite{Lars-EPJA}, it would have been possible to identify the isomeric decay if its lifetime in Th$^{1+}$ was longer than the extraction time ( $\approx$10 ms in continuous extraction).
One possible explanation would be, that the isomer decays already in singly charged $^{229m}$Th via IC, which could only be possible if the energy lies above the second ionization potential of thorium (given with 11.9 eV \cite{secondIonization}). 
Based on the currently accepted value of the isomeric energy 7.8 $\pm$ 0.5 eV \cite{7.8eV}, we consider the IC decay of $^{229m}$Th$^{1+}$ as unlikely.
Also, most pessimistic estimations on the branching ratios involved in the analysis of the underlying data would only allow to shift the expected isomeric energy to a value of 10.5 eV \cite{TkalyaJeet}, which is still below the second ionization potential of thorium. Another explanation would be the depopulation by other decay channels, such as bound-internal conversion, electronic bridge or collisional quenching with buffer gas atoms, that may result in a reduction of the isomeric lifetime in the singly charged state. 

In order to derive the half-life $t_{\mbox{\scriptsize{1/2}}}$, a linear function ($f(t) = -\ln(2)/t_{\mbox{\scriptsize{1/2}}}\cdot t + c$) was fitted to the logarithm of the registered number of counts $N(t)$ in the decay curve. 
The data points were weighted with the inverse of their relative statistical error $1/\sigma_{\mbox{\scriptsize {rel}}} = N/\sqrt{N} = \sqrt{N}$.
Half-lives of $t_{\mbox{\scriptsize{1/2}}}$ = 6.9$\pm$1.0~$\mu$s and $t_{\mbox{\scriptsize{1/2}}}$ = 7.0$\pm$1.0~$\mu$s were obtained for measurements performed with $^{229(m)}$Th$^{2+}$ ions and $^{229(m)}$Th$^{3+}$ ions, respectively.
The corresponding plots and fit functions are shown in Fig.~\ref{Analysis}.
To exclude a distortion of the obtained half-lives from remaining ionic impact signal, a fit-range was chosen whose start point is shifted away $\approx$5 $\mu$s from the falling edge of the ionic impact signal (indicated by the vertical black dashed lines in Fig.~\ref{Analysis}).

The short half-life could potentially explain why some previous experiments aiming on the direct detection of the isomer were not successful.
It is most likely that the isomer decays within the measured 7 $\mu$s via internal conversion, as soon as it is neutralized.
For future experiments, the results can serve as a signature of the isomeric decay and offer a possibility to discriminate between signals generated by ionic impact of $^{229(m)}$Th and its internal conversion decay.
Since the lifetime of the isomer depends on the electronic orbitals in the surrounding of the nucleus, one has to expect that the lifetime is affected by the chemical structure of the surface.
This is taken into account with the error margin of 1.0 $\mu$s.
The influence of the surface on the isomeric lifetime could be probed by coating the micro-channel plate surface with different materials and performing the experiments under UHV conditions to ensure surface cleanliness.
To measure the lifetime of neutral isolated $^{229m}$Th, the ions need to be neutralized without getting into contact with any surface, \textit{e.g.} by charge exchange with caesium vapor in a charge transfer cell as for example described in \cite{CEX}. 

In conclusion, a first measurement of the lifetime of $^{229m}$Th, making use of the internal-conversion decay channel, has been performed, which is a first step towards the full characterization of the isomeric properties.
The inferred half-life is 7.0$\pm$1.0~$\mu$s and well in agreement with theoretical predictions \cite{TkalyaJeet,ElectronicEnvironment2}.
Considering a photonic lifetime of $\tau_\gamma \approx 10^4$ s, as expected for the M1 deexcitation of $^{229m}$Th, the internal conversion coefficient $\alpha_{\text{ICC}} \approx 10^9$, predicted by theory, has been confirmed by our measurements. 
The reported half-life measurements constrain the relative strength of the radiative decay branch of $^{229m}$Th to about $10^{-9}$ in those cases, where IC cannot be significantly suppressed.

We thank S. Stellmer \& T. Schumm for the loan of the Multichannel Scaler and acknowledge fruitful discussions with  S. Stellmer, M. Laatiaoui and J. Crespo L\'{o}pez-Urrutia. 
We also want to thank Christoph Mokry, J\"org Runke, Klaus Eberhardt, Christoph E. D\"ullmann and Norbert G. Trautmann from Johannes Gutenberg University Mainz and Helmholtz Institute Mainz for the production of the uranium sources.
This work was supported by DFG grant (Th956/3-1) and via the European Union's Horizon 2020 research and innovation programme under grant agreement No. 664732 "nuClock".

\end{document}